\documentclass[numberedappendix]{emulateapj}

\newcommand{\himpc}{{\hbox {$~h^{-1}$}{\rm ~Mpc}}}

\slugcomment{Accepted to \textit{The Astrophysical Journal Letters} 02/06/2009}

\shorttitle{GI Correlation Functions of LRGs}
\shortauthors{Okumura \& Jing}

\begin{document}

\title{The Gravitational Shear--Intrinsic Ellipticity Correlation
Functions of \\ Luminous Red Galaxies in Observation and in The
$\Lambda$CDM Model}

\author{Teppei Okumura\altaffilmark{1} and Y.~P. Jing\altaffilmark{1}}

\email{teppei@shao.ac.cn} 

\altaffiltext{1} {Key Laboratory for Research in Galaxies and
Cosmology, Shanghai Astronomical Observatory, Chinese Academy of
Sciences, 80 Nandan RD, Shanghai¡¤200030¡¤China}

\begin{abstract}
We examine whether the gravitational shear--intrinsic ellipticity (GI)
correlation function of the luminous red galaxies (LRGs) can be
modeled with the distribution function of a misalignment angle
advocated recently by Okumura et al. For this purpose, we have
accurately measured the GI correlation for the LRGs in the Data
Release 6 (DR6) of the Sloan Digital Sky Survey (SDSS), which confirms
the results of Hirata et al. who used the DR4 data. By comparing the
GI correlation functions in the simulation and in the observation, we
find that the GI correlation can be modeled in the current
$\Lambda$CDM model if the misalignment follows a Gaussian distribution
with a zero mean and a typical misalignment angle
$\sigma_\theta=34.9^{+1.9}_{-2.1}$ degrees. We also find a correlation
between the axis ratios and intrinsic alignments of LRGs. This effect
should be taken into account in theoretical modeling of the GI and
intrinsic ellipticity--ellipticity correlations for weak lensing
surveys.
\end{abstract}


\section{Introduction}
Weak gravitational lensing by large-scale structure is a promising
tool to directly probe matter distribution in the universe.  The main
source of contaminations for weak lensing observations comes from two
types of intrinsic alignments: the ellipticity correlation of source
galaxies with each other [intrinsic ellipticity--ellipticity (II)
correlation] and the ellipticity correlation of lense galaxies with
the surrounding matter distribution [gravitational shear--intrinsic
ellipticity (GI) correlation].  Investigating these intrinsic
alignments is also important for galaxy formation studies.

There has been much theoretical work \citep[e.g.,][]{Heavens2000,
Croft2000, Lee2000, Catelan2001,Jing2002} as well as observational
work \citep[e.g.,][]{Pen2000, Brown2002, Hirata2004b, Heymans2004,
Mandelbaum2006, Okumura2009} which attempted to estimate the II
correlation. In \citet{Okumura2009}, we have determined the II
correlation of luminous red galaxies (LRGs) accurately to a large
scale ($\sim 30 \himpc$) using the Data Release 6 of the Sloan Digital
Sky Survey \citep[SDSS DR6;][]{Y2000}. From this measurement, we
further gave a constraint on the misalignment angle of typically
$35^{\circ}$ ($26^{\circ}$ on average) between the central giant
elliptical galaxies and their host dark matter halos using the halo
occupation distribution (HOD) approach
\citep[e.g.,][]{Jing1998,Zheng2008}. This misalignment is found to be
in agreement with the misalignment between the inner dark matter and
the host halos by \citet{Faltenbacher2009} using an $N$-body
simulation, though it is unknown if this is a coincidence because the
light distribution of LRGs is not necessarily to be the same as the
inner dark matter distribution of halos. Nevertheless, the
observational constraint does provide very useful clues to
understanding the contaminations on weak lensing surveys as well as to
studying formation of giant ellipticals \citep{Okumura2009}.

Given the distribution of the misalignment angles, we can also predict
the GI of LRGs with the surrounding dark matter distribution for
current cosmological models by combining an $N$-body simulation and
the HOD modeling. Such predictions can be compared with the
observations of the GI such as those of \citet{Hirata2007} and
\citet{Faltenbacher2009}. The comparisons will further test if the
distribution of the misalignment angle is correct, and will
furthermore tell us how to model the GI contaminations generally for
future weak lensing surveys. This approach is complementary to
recently proposed statistical approaches which attempt to eliminate or
minimize the GI effect in weak lensing surveys purely based on
observational samples \citep{Hirata2004a, King2005, Heymans2006,
Bridle2007, Joachimi2008, Zhang2008, Hui2008}.  We will present such a
study in this {\it Letter}.

\section{SDSS Luminous Red Galaxy Sample}\label{sec:sdss}
Our HOD model parameters used in Section \ref{sec:model} are from
\citet{Seo2008} which is based on the fitting of \citet{Zheng2008} to
the projected correlation function of LRGs of \citet{Zehavi2005}. When
we started the work, we tried to use the observational data of GI of
\citet{Hirata2007}, but later we found that the projected correlation
function presented in their paper is slightly smaller (about 20 \% in
low-$z$ and 15 \% in high-$z$ subsamples) than that in
\citet{Zehavi2005}. While we do not know if the samples they used are
slightly different, it is apparent that the HOD parameters of
\citet{Seo2008} cannot be applied to the sample of
\citet{Hirata2007}. Therefore, we decided to measure the GI of LRGs
using the SDSS DR6 data \citep{Y2000, E2001,
Adelman-McCarthy2008}. The LRG sample is the same as that used in our
previous paper \citep{Okumura2009}. There are 83,773 LRGs in the
redshift range $0.16<z<0.47$.

Here we need to model the radial and angular selection functions of
the observational sample.  We build the radial selection function by a
spline fit with a Gaussian smoothing of the redshift distribution of
observed LRGs. The angular selection function is constructed using the
angular survey mask provided by the Value Added Galaxy Catalog
\citep[VAGC;][]{Blanton2005}. We assumed that our sample has the same
survey geometry as that of the VAGC and excluded all the LRGs which
are not overlapped. Then the number of LRGs is reduced to
78,758. Finally we identify 73,935 central LRGs using the criteria
proposed by \citet{Reid2008}.

\begin{figure}[bt]
\epsscale{1.00} \plotone{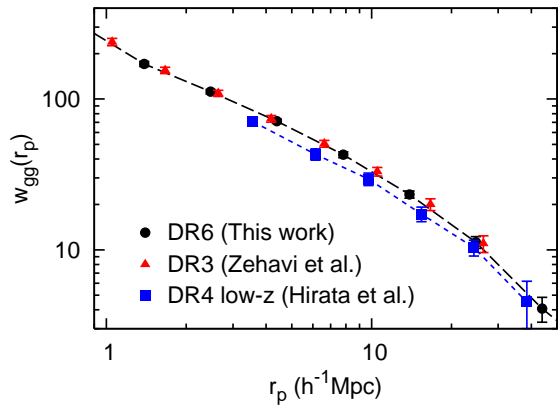}
\caption{Projected galaxy-galaxy correlation function for the LRG
sample at $0.16<z<0.36$. For comparison, the data of
\citet{Zehavi2005} and \citet{Hirata2007} are also plotted.  \\~ }
\label{fig:gg}
\end{figure}

We measure the projected correlation function from the LRG sample,
$w_{gg}(r_p)=\int\xi_{gg}(r_p,\Pi)d\Pi$, where $\xi_{gg}(r_p,\Pi)$ is
the galaxy autocorrelation function as a function of separations
perpendicular ($r_p$) and parallel ($\Pi$) to the line of sight and is
measured using the \citet{LS1993} estimator. Figure \ref{fig:gg} shows
the comparison of our measured $w_{gg}$ with the previous work by
\citet{Zehavi2005}.  Very good agreement between the two studies
confirmed that our selection functions work well on the scales of
interests in this work.

To measure the GI function of LRGs, we also need information on their
shapes.  The ellipticity of galaxies is determined by the SDSS
photometric pipeline called {\tt photo} and defined as the ellipticity
of the 25 mag arcsec${}^{-2}$ isophote in the $r$ band
\citep{Stoughton2002}. The point-spread function has been corrected
when measuring the shapes \citep{Lupton2001}. The components of the
ellipticity are defined as
\begin{eqnarray}
  \left(
  \begin{array}{c} e_+ \\ e_\times \end{array}
  \right) = \frac{1-q^2}{1+q^2} \left(
  \begin{array}{c}\cos{2\beta} \\ \sin{2\beta}\end{array}
  \right), \label{eq:ellip}
\end{eqnarray}
where $q$ is the ratio of minor and major axes and $\beta$ is the
position angle of the ellipticity from the north celestial pole to
east.

\section{Gravitational Shear--Intrinsic Ellipticity Correlation of LRGs}
\label{sec:gi}
To estimate the GI correlation, we adopt the formalism developed by
\citet{Mandelbaum2006} and \citet{Hirata2007}.  The generalized
\citet{LS1993} estimator is used for estimating the GI correlation
function,
\begin{equation}
  \xi_{g+}(r_p, \Pi) = \frac{S_+(D-R)}{RR} 
\end{equation}
where $RR$ is the normalized counts of random--random pairs in a
particular bin in the space of $(r_p,\Pi)$. $S_+D$ is the sum over all
pairs of the $+$ component of shear in $(r_p,\Pi)$,
\begin{equation}
  S_{+}D = \sum_{i\neq j | r_p,\Pi} \frac{e_{+}(j|i)}{2 {\cal R}}, 
  \label{eq:S+D}
\end{equation}
where the ellipticity component of $j$th LRG, $e_{+}(j|i)$, is
redefined relative to the direction to the $i$th LRG and thus
corresponds to the elongation along the direction, and ${\cal
R}=1-\sigma_{\rm SN}^2=1-\langle e_+^2\rangle$ is the shear
responsivity \citep[e.g.,][]{Bernstein2002} and ${\cal R}\approx
0.947$ for our LRG sample. $S_+R$ is calculated likewise using a
catalog of randomly distributed points in the survey region.  Finally
the projected GI correlation function is obtained by doing the
projection along the radial direction,
\begin{equation}
  w_{g+}(r_p)=\int^{\Pi_{\rm max}}_{-\Pi_{\rm max}}
  \xi_{g+}(r_p,\Pi)d\Pi. \label{eq:gi}
\end{equation}
We adopt $\Pi_{\rm max}=80\himpc$, but changing $\Pi_{\rm max}$ from
60 to $100\himpc$ does not significantly change $w_{g+}(r_p)$ for
$r_p<100\himpc$.  Positive $w_{g+}(r_p)$ means that the major axes of
LRGs tend to point toward overdensities at a transverse scale $r_p$.
$w_{g+}$ is related to the density--intrinsic ellipticity correlation
function $w_{\delta +}$ through the galaxy bias $b_g$ as $w_{g+}=b_g
w_{\delta +}$ on large scales \citep[see also Section
\ref{sec:discussion}]{Mandelbaum2006, Hirata2007}.
$w_{g\times}(r_p)$ is also calculated in the same way and used for a
test of systematics because $w_{g\times}(r_p)$ should be zero on all
scales.

\begin{figure}[bt]
\epsscale{1.00}
\plotone{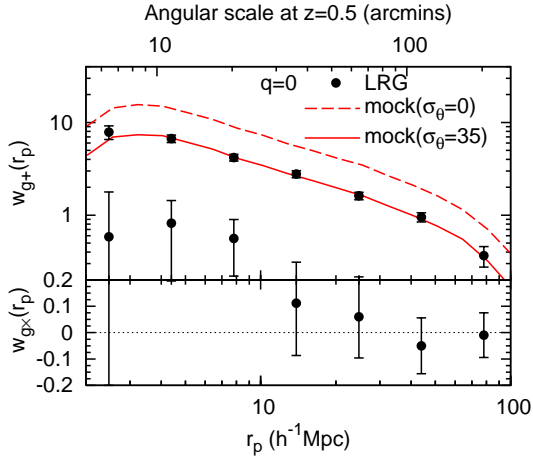}
\caption{Projected GI correlation functions from the SDSS central
LRGs, $w_{g+}(r_p)$ and $w_{g\times}(r_p)$. The dashed and solid lines
are respectively the model predictions for $w_{g+}$ obtained from the
mock LRGs with $\sigma_\theta=0^{\circ}$ and $35^{\circ}$. Mixed
logarithmic and linear scalings are used for the vertical axis. The
horizontal axis at the top shows the corresponding angular scale when
the transverse separations are located at $z=0.5$.  \\~ }
\label{fig:gi}
\end{figure}

The resulting GI correlation functions for the observed central LRGs
are shown in Figure \ref{fig:gi}. The error bars represent $1\sigma$
errors estimated from 93 jackknifed realizations. Here the axis ratio
$q$ in equation (\ref{eq:ellip}) is set to be zero. This is equivalent
to the assumption that a galaxy is a line along its major axis.  In
this case the shear responsivity becomes ${\cal R}=1-\langle
\cos^2{2\beta}\rangle= 0.5$. Clear detection of the GI correlation
$w_{g+}$ can be seen up to large scales.  Note that when our result is
compared with the previous work of \citet{Hirata2007}, the difference
of the $q$ factor and thus of ${\cal R}$ must be taken into account.
The prescription of $q=0$ and its relationship with that of $q>0$ will
be discussed in Section \ref{sec:orientation}. From the plots of
$w_{g\times}$ in Figure \ref{fig:gi} we confirm that all the values
are consistent with zero at the $2\sigma$ confidence level.  We thus
discuss the GI correlation only in terms of $w_{g+}$ in the following
analysis.

\section{Comparison to Model Predictions} \label{sec:model}

\subsection{Modeled Gravitational Shear--Intrinsic Ellipticity Correlation Function}\label{sec:hecf}

To make model predictions for the GI correlation, we follow the same
methodology as in \citet{Okumura2009}.  We use a halo catalog
constructed from a high-resolution cosmological simulation with
$1024^3$ particles in a cubic box of side $1200\himpc$
\citep{Jing2007}.  Central galaxies are assigned to the simulated
halos using the best-fit HOD parameters for LRGs found by
\citet{Seo2008}\citep[see also][]{Zheng2008}.  The resulting fraction
of mock central LRGs is 93.7\% and we use only the centrals in order
to compare with our observation.

We consider halos to have triaxial shapes \citep{JS2002}. The two
components of the ellipticity of each halo are estimated from the
second moments of the projected mass distribution
\citep[e.g.,][]{Croft2000}
\begin{equation}
  \left(
  \begin{array}{c} e_+ \\ e_\times \end{array}
  \right) = \frac{1}{I_{xx}+I_{yy}} \left(
  \begin{array}{c}I_{xx}-I_{yy} \\ 2I_{xy}\end{array}
  \right), \label{eq:ellip_moment}
\end{equation}
where $I_{ij}=\frac{1}{N}\sum^N x_i x_j$ and $N$ is the number of
particles in a halo.  Then the GI correlation function of halos is
measured in the same way as that of LRGs, where the value of $q$ is
assumed to be zero again and thus ${\cal R}=0.5$.

First we assume that all central galaxies are completely aligned with
their parent dark matter halos. The GI correlation function of central
galaxies is then calculated and shown in Figure \ref{fig:gi}. In order
to refine the statistics, we averaged over seven mock samples with
different random seeds for assigning LRGs to dark halos.
Interestingly, the GI correlation function of the mock LRGs, when they
are assumed to be aligned completely with their host halos, has the
same shape as but is about twice as high as the observation. Similar
results were shown for the II correlation \citep{Okumura2009}.

 \subsection{Constraints on Misalignment}\label{sec:misalignment}

In this subsection, we consider the case in which the major axis of
each central galaxy is misaligned with that of its host halo and give
a constraint on the misalignment angle by comparing the observed GI
correlation function $w_{g+}$ with its model prediction. Following
\citet{Okumura2009} we assume that the misalignment angle $\theta$
between the major axes of central LRGs and their host halos follows a
Gaussian function with a zero mean and a width $\sigma_\theta$, where
$\sigma_\theta$ is the typical misalignment angle. We artificially
assign misalignment to the position angle of each mock central LRG
relative to its host halo according to the Gaussian function.  For
each chosen value of $\sigma_\theta$ and each LRG mock sample, we
generate nine misaligned LRG samples by choosing different random
seeds.  Our model prediction for each $\sigma_\theta$ is thus
calculated by averaging over $7 \times 9=63$ misaligned samples.

In comparing the observational data with the model prediction,
$\chi^2$ statistics are calculated in the range of
$20^{\circ}<\sigma_\theta<50^{\circ}$.  In this analysis we use the
seven data points of $w_{g+}(r_p)$ shown in Figure \ref{fig:gi}, while
there is one free parameter, $\sigma_\theta$; thus the degree of
freedom is 6. The covariance matrix estimated using 93 jackknifed
subsamples is used for the calculation of $\chi^2$.

The fits of the observed GI correlation function $w_{g+}$ to the model
prediction give a tight constraint on the misalignment parameter,
$\sigma_\theta = 34.9^{+1.9}_{-2.1}$ degrees (68\% confidence level),
which corresponds to a mean misalignment angle of $26^{\circ}.9$.
This is in very good agreement with our previous work on the II
correlation which gave $\sigma_\theta = 35.4^{+4.0}_{-3.3}$ degrees.
The constraint from the GI correlation is tighter than that from the
II correlation because the GI correlation is better determined. The
model prediction of $w_{g+}$ with $\sigma_\theta=35^{\circ}$ is shown
in Figure \ref{fig:gi}.

\subsection{Correlation of the LRG shape and its orientation}
\label{sec:orientation}
The misalignment parameter was constrained with the assumption of
$q=0$, i.e., we considered the orientation of the LRGs relative to
their spatial distribution only. If there is no correlation between
the shape of the LRGs and their orientation, we can use the
misalignment angle distribution to model the GI of LRGs even when the
shape is included \citep[i.e., GI in weak lensing studies;][]
{Hirata2007}. In order to see if this correlation exists, we define a
normalized GI correlation function $\bar{w}_{g+}$ as
\begin{equation}
  \bar{w}_{g+}(r_p;q)=\left\langle{\frac{1-q^2}{1+
  q^2}}\right\rangle^{-1} w_{g+}(r_p;q), \label{eq:giq}
\end{equation}
where $\left\langle{\frac{1-q^2}{1+ q^2}}\right\rangle$ is the value
averaged over all objects in the sample (it is $0.29$ for observed
LRGs and $0.42$ for mock host halos of LRGs), and $w_{g+}(r_p;q)$ is
the same as equation (\ref{eq:gi}) except the $q$ dependence is
included. If there is no correlation between axis ratios and
orientations, we expect $\bar{w}_{g+}(r_p;q=q_{\rm
LRG})={w}_{g+}(r_p;q=0)$ for observed LRGs and
$\bar{w}_{g+}(r_p;q=q_{\rm mock})={w}_{g+}(r_p;q=0)$ for halos. Here
we neglect the factor of the shear responsivity, $1/2{\cal R}$, so
equation (\ref{eq:giq}) just corresponds to the replacement of
$1/2{\cal R}$ by $\left\langle{\frac{1-q^2}{1+
q^2}}\right\rangle^{-1}$ in equation (\ref{eq:S+D}).
 
\begin{figure}[bt]
\epsscale{1.00}
\plotone{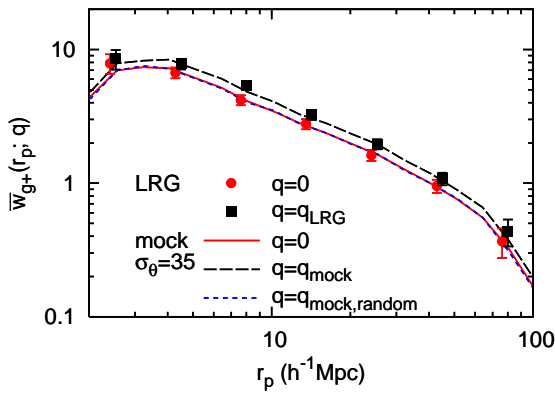}
\caption{Normalized GI correlation functions $\bar{w}_{g+}(r_p;q)$.
All the lines are the model predictions with
$\sigma_\theta=35^{\circ}$.  The dotted blue line is the one with
$q_{\rm mock}$ shuffled randomly.  The circles/squares have been
respectively offset in the horizontally negative/positive direction
for clarity.  \\~ }
\label{fig:q}
\end{figure}

In Figure \ref{fig:q} we show $\bar{w}_{g+}(r_p;0)$ of the observed
LRGs and mock LRGs with $\sigma_\theta = 35^{\circ}$, which are the
same data as those in Figure \ref{fig:gi} because
$\bar{w}_{g+}(r_p;0)=w_{g+}(r_p;0)$. Next we calculate
$\bar{w}_{g+}(r_p;q_{\rm mock})$ for the mock LRGs with $q_{\rm mock}$
determined from the dark matter distribution within the host halos,
which is shown as the dashed line in Figure \ref{fig:q}.  These
results indicate that there exists a correlation between the shape and
orientation of the dark matter halos, and this correlation leads to an
increase of $\sim 15 \%$ in the normalized GI correlation. If the
shapes of the halos were uncorrelated with their orientations, we
expect that this $\bar{w}_{g+}(r_p;q_{\rm mock})$ would be equal to
$\bar{w}_{g+}(r_p;0)$. This is indeed confirmed in the figure, where
the dotted line, the correlation when the values $q_{\rm mock}$ (not
the orientation) of host halos are shuffled randomly, is completely
overlapped with the solid line.  Almost the same amount of increase is
seen for the GIs of the observed LRGs, as shown by the squares and the
circles in the figure, which implies that there is a correlation of
the shapes and orientation in the observation.
 
This correlation should be taken into account when one models the GI
and II effects in theory by using the distribution function of the
misalignment angle. It should also be considered if
$w_{g+}(r_p;q=q_{\rm LRG})$ is used to constrain the distribution
function of the misalignment angle. If the correlation is neglected,
the misalignment angle could be slightly underestimated. This is the
reason why only the orientation of the LRGs are used in
\citet{Okumura2009} and in the current work to constrain the
misalignment angle.

\section{Discussion}\label{sec:discussion}
In this Letter, we examined whether the GI correlation of LRGs can be
modeled with the distribution function of misalignment angle advocated
by \citet{Okumura2009} based on the II correlation. For this purpose,
we have accurately measured the GI correlation for the LRGs in the
SDSS DR6, which also confirms the results of \citet{Hirata2007} who
used the DR4 data. By comparing the GI correlations in the simulation
and in the observation, we found that the GI correlation can be
modeled in the current $\Lambda$CDM model, if the misalignment follows
a Gaussian distribution with a zero mean and a typical misalignment
angle $\sigma_\theta=34.9^{+1.9}_{-2.1}$ degrees. This constraint on
$\sigma_\theta$ is in excellent agreement with the previous work,
$\sigma_\theta=35.4^{+4.0}_{-3.3}$ degrees, based on the II
correlation. The constraint on $\sigma_\theta$ is tighter in this
Letter, because the GI correlation is better determined than the II
correlation in the observation. Furthermore, the good agreement of the
observed and theoretical GI functions further lends nontrivial
support for the $\Lambda$CDM scenarios and for the distribution
function of the misalignment angle.

We have found a correlation between the axis ratios and intrinsic
alignments of LRGs. If the correlation is neglected, one would
underestimate the GI correlation (in the case of $q\ne 0$) by $\sim 15
\%$ if the shape $q$ and orientation $\theta$ of the LRGs are randomly
chosen from their distribution functions. This effect should be taken
into account in theoretical modeling of the GI and II correlations for
weak lensing surveys.

These results have profound implications both for future weak lensing
surveys and for studying the formation of giant elliptical galaxies.
For weak lensing surveys, the relevant quantity is the correlation
function between the mass overdensity and the intrinsic ellipticity,
$w_{\delta +}$.  As \citet{Mandelbaum2006} and \citet{Hirata2007}
reasoned, these two quantities are simply related through the galaxy
bias as $w_{g+}= b_g w_{\delta +}$. We can also measure $w_{\delta +}$
directly in our simulation. Our result supports that $b_g$ in this
simple relation is about 2 for all the scales explored here and for
our chosen $\Lambda$CDM model.

\acknowledgments 

This work is supported by NSFC (10533030, 10821302, 10878001), by the
Knowledge Innovation Program of CAS (KJCX2-YW-T05), and by 973 Program
(2007CB815402). Numerical calculations are in part performed on a
parallel computing system at Nagoya University.  Funding for the SDSS
and SDSS-II has been provided by the Alfred P. Sloan Foundation, the
Participating Institutions, the National Science Foundation, the
U.S. Department of Energy, the National Aeronautics and Space
Administration, the Japanese Monbukagakusho, the Max Planck Society,
and the Higher Education Funding Council for England. The SDSS web
site is http://www.sdss.org.

\end{document}